\documentclass[
aps, prx,
 amsmath,amssymb,
reprint,%
floatfix, superscriptaddress]{revtex4-2}
\usepackage{booktabs}
\usepackage{hhline}

\usepackage[version=4]{mhchem}
\usepackage{siunitx}
\DeclareSIUnit\angstrom{\text{\AA}}

\usepackage{svg}
\usepackage{graphicx}
\usepackage{dcolumn}
\usepackage{bm}

\usepackage{hyperref}
\hypersetup{
    unicode=false,          
    pdftoolbar=true,        
    pdfmenubar=true,        
    pdffitwindow=false,     
    pdfstartview={FitH},    
    colorlinks=true,        
    linkcolor=blue,         
    citecolor=blue,         
    urlcolor=blue           
}

\usepackage{mathptmx}
\usepackage{etoolbox}

\usepackage{float}

\pdfoutput=1

\begin{document}

\title{\textcolor{blue}{Publisher's version available at https://doi.org/10.1002/adfm.202308679}\\Toward Sustainable Ultrawide Bandgap van der Waals Materials: \\An ab initio Screening Effort}

\author{Chuin Wei Tan}
\affiliation{Science, Mathematics and Technology (SMT) Cluster, Singapore University of Technology and Design, 8 Somapah Rd, Singapore 487372}
\affiliation{John A.Paulson School of Engineering and Applied Sciences, Harvard University, Cambridge, Massachusetts 02138, United States}

\author{Linqiang Xu}
\affiliation{Science, Mathematics and Technology (SMT) Cluster, Singapore University of Technology and Design, 8 Somapah Rd, Singapore 487372}
\affiliation{State Key Laboratory of Mesoscopic Physics and Department of Physics, Peking University, Beijing 100871, P. R. China}

\author{Chen Chen Er}
\affiliation{Science, Mathematics and Technology (SMT) Cluster, Singapore University of Technology and Design, 8 Somapah Rd, Singapore 487372}

\author{Siang-Piao Chai}
\affiliation{Multidisciplinary Platform of Advanced Engineering, Department of Chemical Engineering, School of Engineering, Monash University Malaysia, Jalan Lagoon Selatan, 47500 Bandar Sunway, Selangor, Malaysia}

\author{\\Boris Kozinsky}
\affiliation{John A.Paulson School of Engineering and Applied Sciences, Harvard University, Cambridge, Massachusetts 02138, United States}
\affiliation{Robert Bosch LLC Research and Technology Center, Watertown, Massachusetts 02472, United States}

\author{Hui Ying Yang}
\affiliation{Engineering Product Development (EPD), Singapore University of Technology and Design, 8 Somapah Rd, Singapore 487372}

\author{Shengyuan A. Yang}
\affiliation{Science, Mathematics and Technology (SMT) Cluster, Singapore University of Technology and Design, 8 Somapah Rd, Singapore 487372}

\author{Jing Lu}
\affiliation{State Key Laboratory of Mesoscopic Physics and Department of Physics, Peking University, Beijing 100871, P. R. China}
\affiliation{Collaborative Innovation Center of Quantum Matter, Beijing 100871, P. R. China}
\affiliation{Beijing Key Laboratory for Magnetoelectric Materials and Devices, Beijing 100871, P. R. China}
\affiliation{Peking University Yangtze Delta Institute of Optoelectronics, Nantong 226000, P. R. China}
\affiliation{Key Laboratory for the Physics and Chemistry of Nanodevices, Peking University, Beijing 100871, P. R. China}

\author{Yee Sin Ang}
\email{yeesin\_ang@sutd.edu.sg}
\affiliation{Science, Mathematics and Technology (SMT) Cluster, Singapore University of Technology and Design, 8 Somapah Rd, Singapore 487372}

\begin{abstract}
The sustainable development of next-generation device technology is paramount in the face of climate change and the looming energy crisis. Tremendous effort has been made in the discovery and design of nanomaterials that achieve \emph{device-level} sustainability, where high performance and low operational energy cost are prioritized. However, many of such materials are composed of elements that are under threat of depletion and pose elevated risks to the environment and human health. The role of \emph{materials-level} sustainability in computational screening efforts has been overlooked thus far. This work presents a general van der Waals materials screening framework imbued with sustainability-motivated search criteria. Using ultrawide bandgap (UWBG) materials as a backdrop, 25 sustainable UWBG layered materials comprising only of low-risks elements resulted from this screening effort, with several meeting the requirements for dielectric, power electronics, and ultraviolet device applications. These findings constitute a critical first-step towards reinventing a more sustainable electronics landscape beyond silicon, with the framework established in this work serving as a harbinger of sustainable 2D materials discovery.
\end{abstract}

\maketitle

\section{\label{Introduction} Introduction}

Since the isolation of graphene in 2004 \citep{graphene_isolation}, two-dimensional (2D) layered materials have received immense interest for their myriad of unusual physical properties. Of significance to the development of next-generation nanoelectronics is the family of 2D semiconductors, including those derived from van der Waals (vdW) layered bulk parents, such as transition metal dichalcogenides (TMDCs) \citep{2DTMDC}, and those synthetically derived, such as \ce{MA2Z4} \citep{MA2Z4_1,MA2Z4_2,MA2Z4_3}, Janus TMDCs \citep{janus_TMDC} and liquid-metal-printed ultrathin oxides \citep{2D_TeO2}. By virtue of their ultrathin bodies and excellent electrical properties \citep{2D_FET}, 2D semiconductors form an emerging materials platform to realize sub-10~\si{nm} gate length field-effect transistors (FETs) \citep{sub10_2DFETs} -- an otherwise insurmountable challenge using conventional silicon-based device technology. 
Recent demonstrations of ultralow contact resistance below 100~\si{\Omega \mu m} \citep{cont_res_1, cont_res_2, cont_res_3}, steep-slope devices \citep{slope_1, slope_2, slope_3}, ultrascaled transistors \citep{ultrascale_1, ultrascale_2}, and chip-scale integration \citep{chip1, chip2, chip3} have cemented the prospects of 2D semiconductors as a building block in low-power, high-performance transistors \citep{2D_transistors_1, 2D_transistors_2} critical for sustainable computing electronics. These achievements are based largely on 2D semiconductors with moderate bandgaps, leaving the subgroup of ultrawide bandgap (UWBG) semiconductors relatively overlooked. Conventional UWBG semiconductors, such as SiC and GaN, hold important roles in dielectric applications, power electronics and deep-ultraviolet photonics \citep{UWBG_review}. The integration of 2D materials in UWBG semiconductor device technology thus presents new opportunities that warrant further exploration.

With the development of highly-parallel density functional theory software, modern supercomputing systems and large materials databases \citep{materials_project, materials_cloud, 2dmatpedia, c2db1, c2db2}, computational screening has emerged as an effective strategy in accelerating materials discovery and design. Previous 2D materials screening efforts have covered diverse functions ranging from photocathode coatings \citep{ht_1} and thermoelectrics \citep{ht_2} to piezoelectrics \citep{ht_3}, ferroelectrics \citep{ht_4} and oxygen electrocatalysts \citep{ht_5}. While functionality and material stability form the main thrusts of these screening endeavors, sustainability concerns are often neglected. Even when the notion of sustainability is invoked, it refers to \emph{device-level} features rather than \emph{materials-level} sustainability. Take for example the recently proposed UWBG layered dielectrics \ce{Bi2SeO5}, an experimentally synthesized native oxide related to high-mobility \ce{Bi2O2Se} \citep{Bi2SeO5_1, Bi2SeO5_2}, and \ce{LaOCl}, the result of a dielectric screening effort \citep{2d_dielectric_screen}. Their high dielectric constants ($\kappa$) and low leakage currents make them excellent gate dielectric materials capable of reducing power wastage and optimizing device efficiency, thus improving the sustainability of FETs at the \emph{device level}. Their comprising elements, however, are not. The known reserves of \ce{Bi} and \ce{Se} are expected to be depleted within 50 years \citep{sus_chemsuschem} and \ce{La} is a rare Earth element whose mining has adversely impacted the environment and human health \citep{rare_earth_elements}. Ostensibly, the inclusion of stringent materials-sustainability-focused screening criteria could significantly diminish the pool of high-performing candidates identified for specific applications. These considerations lead to the central questions of this work: How sizable would the pool of sustainable UWBG 2D semiconductors be, and to what extent can these candidates meet specific (opto)electronic device requirements given the imposition of such selective sustainability-centric screening filters?

In this work, we screen the 2DMatPedia \citep{2dmatpedia} database for likely synthesizable UWBG 2D semiconductors composed of low-risk elements that may improve the sustainability of 2D semiconductor device technology at the `upstream' materials level. The sustainability-focused screening strategy introduced in this work filters out elements that pose elevated risks to the environment and human health, and are themselves at risk of depletion. The candidates are subsequently characterized with ab initio calculations, largely based on density functional theory (DFT), and assessed for their performance in various applications. Despite the stringent sustainability-motivated search criteria, we identify multiple candidates with promising performance as (i) high-$\kappa$ gate dielectrics compatible with low standby power operation, (ii) low-$\kappa$ spacer materials with good mechanical strength, and (iii) broadband deep ultraviolet (UV) photodetectors and polarizers. Using ab initio quantum transport simulations, we further demonstrate that \ce{CuI}, one of the screened candidates, can function as a channel material in sub-10~\si{nm} FETs to fulfill the operation requirements of the International Technology Roadmap for Semiconductors (ITRS) \citep{ITRS}, even at elevated temperatures up to 600~\si{K}.

Our findings constitute a critical first-step towards reinventing a more sustainable electronics landscape beyond silicon. The focus on \emph{materials-level} sustainability in this work presents a radically different perspective on 2D materials screening efforts. The framework established here shall serve as a harbinger of sustainable 2D materials discovery, and can be generalized for screening other classes of sustainable 2D functional materials.

\section{Results and Discussion}

\subsection{Screening Procedure and Summary of Candidates}

\begin{figure*}[ht]
\centering
\includegraphics[width=0.95\linewidth]{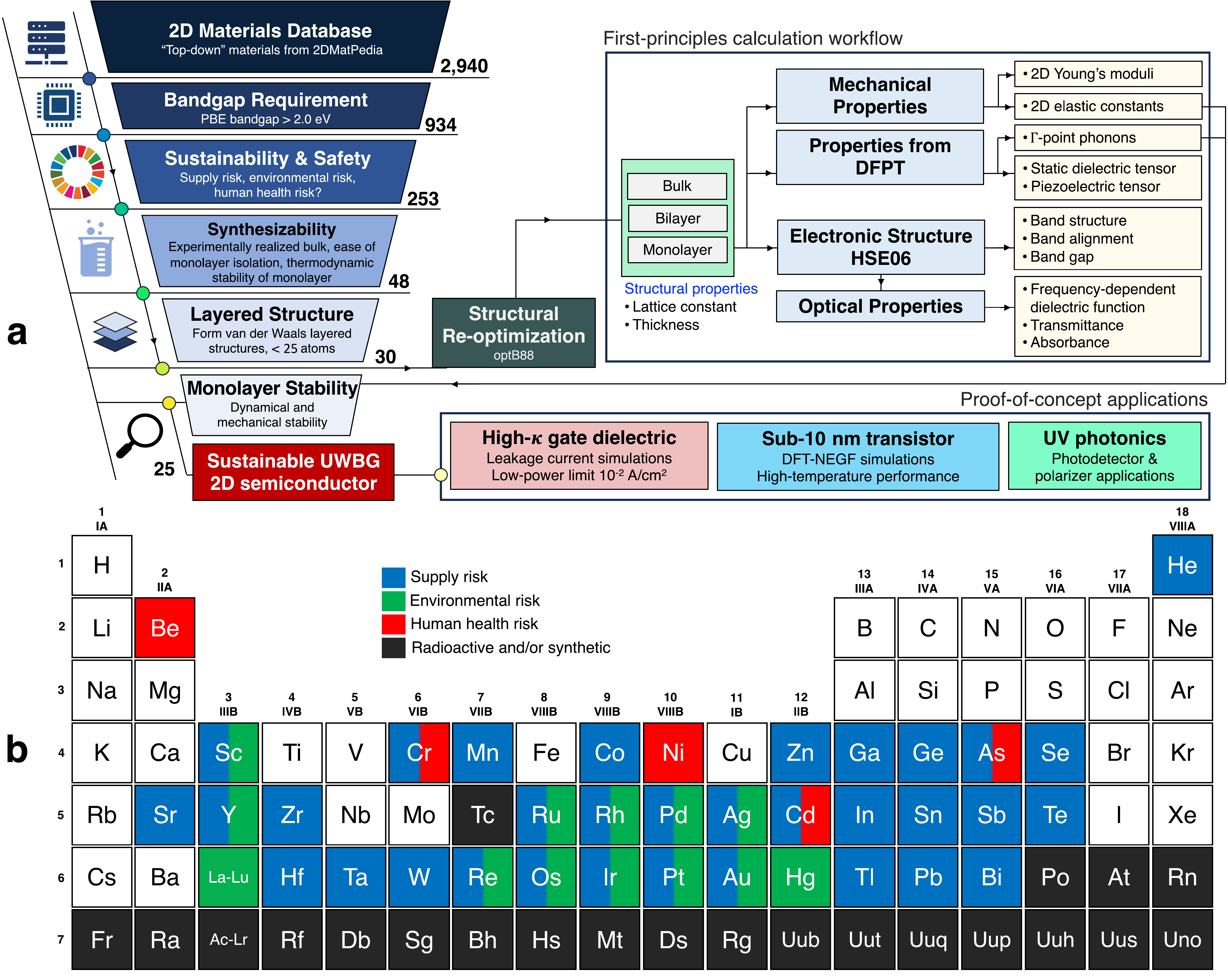}
\caption{\textbf{Computational screening process.} (a) Overview of the computational screening procedure and calculations performed to assess each candidate's performance in various ultrawide bandgap applications. (b) Periodic table of low-risk elements used for the \emph{sustainability and safety} screening criteria.}
\label{overview}
\end{figure*} 

In this section, we first introduce a broadly applicable sustainability-focused 2D materials screening framework that can also be tailored for other applications. We then discuss how this framework is employed to catalogue sustainable 2D UWBG semiconductors that are well suited for dielectric, transistor and UV optical device applications.

Our screening procedure is based on three guiding principles - \emph{functionality} (ultrawide bandgap), \emph{sustainability and safety}, as well as \emph{synthesizability and stability}. An overview of the screening procedure and the calculations performed in this work is presented in Figure~\ref{overview}a. \\

\textbf{Functionality}. The key functional attribute considered in this work is an ultrawide bandgap, which we define as a gap exceeding 3~\si{eV} \citep{2d_wide_gap}. As 2DMatPedia only provides PBE gaps, well-known to underestimate experimentally measured bandgaps, we screen for PBE gaps exceeding 2~\si{eV}, which map onto more accurate HSE06 gaps above 3~\si{eV} \citep{monolayer_bandgap_benchmark}. \\

\textbf{Sustainability and safety}. The aim of this filter is to select materials composed of low-risk elements, as presented in the periodic table of Figure~\ref{overview}b, based on the following risk factors. (i) \emph{Supply risk}. The periodic tables of `endangered elements' from References~\onlinecite{sus_pnas}, \onlinecite{sus_acs}, \onlinecite{sus_euchemsoc} and \onlinecite{sus_chemsuschem} are analyzed and combined to form an element-based filter for excluding candidate structures whose constituting elements are under serious supply risk as explained in the Supporting Information (Section~S1). (ii) \emph{Environmental risk}. We exclude environmentally damaging elements identified in the criticality study of Reference~\onlinecite{sus_pnas} with criticality scores exceeding 50\%. This study found that gold and platinum group noble metals were the most harmful due to their extraction and processing methods. The rare Earth elements, including all 15 lanthanides, \ce{Y} and \ce{Sc}, are also excluded due to the environmental harm associated with mining them \citep{rare_earth_elements}. (iii) \emph{Human health risk}. We additionally exclude elements known to be carcinogenic, namely \ce{As}, \ce{Be}, \ce{Cd}, \ce{Ni} and \ce{Cr}, as declared by the International Agency for Research on Cancer \citep{IARC}. \textcolor{black}{In fact, \ce{Pb}, \ce{Cd}, \ce{Hg} and hexavalent \ce{Cr} are restricted under the European Union's Restriction of Hazardous Substances in Electrical and Electronic Equipment (RoHS) directive \citep{rohs}.} Humans are exposed to these elements through mining and industry \citep{element_cancer}, such as in the case of \ce{Cd}, a widespread industrial pollutant emitted to the air by mines, metal smelters and industrial facilities \citep{cadmium}. Another avenue for exposure is improper e-waste disposal, a problem exacerbated by the poor e-waste recycling rates below 20\% \citep{ewaste_2} in spite of its volume growing at 3 to 5\% annually \citep{ewaste_1}. \\

\textbf{Synthesizability and stability}. \textcolor{black}{Since the initial exfoliation of graphene with the `Scotch tape' method, top-down exfoliation-based methods are still viewed as inexpensive and more accessible alternatives to bottom-up methods such as chemical vapor deposition in synthesizing ultrathin 2D materials \citep{exfol_review_1}. Modern exfoliation techniques such as shear exfoliation, ball milling and electrochemical intercalation are capable to producing few-layered 2D materials spanning 100~\si{\mu m} wide, with the potential for producing bulk quantities \citep{exfol_review_1, exfol_review_2}. We hence focus on exfoliation-based synthesis in this work and only consider monolayers that are likely to be exfoliated from already experimentally found bulk parents.} Methodologically, this translates to only considering the `top-down' structures from 2DMatPedia, that is, the ones theoretically exfoliated from their bulk parents in the Materials Project database \citep{materials_project}. We further require the bulk parents to have an experimental record based on Material Project's matching algorithm to entries in the Inorganic Crystal Structure Database (ICSD), and lie within 10~\si{meV}/atom of their convex hulls in Materials Project. Additionally, an upper bound of 100 \si{meV}/atom was imposed for the exfoliation energies provided by 2DMatPedia to increase the likelihood of exfoliation. As for stability, we only select structures with decomposition energies below 100~\si{meV}/atom since 53 experimentally synthesized structures found in 2DMatPedia fulfil this requirement \citep{2dmatpedia}. \\

\begin{figure*}[htb]
\centering
\includegraphics[width=0.95\linewidth]{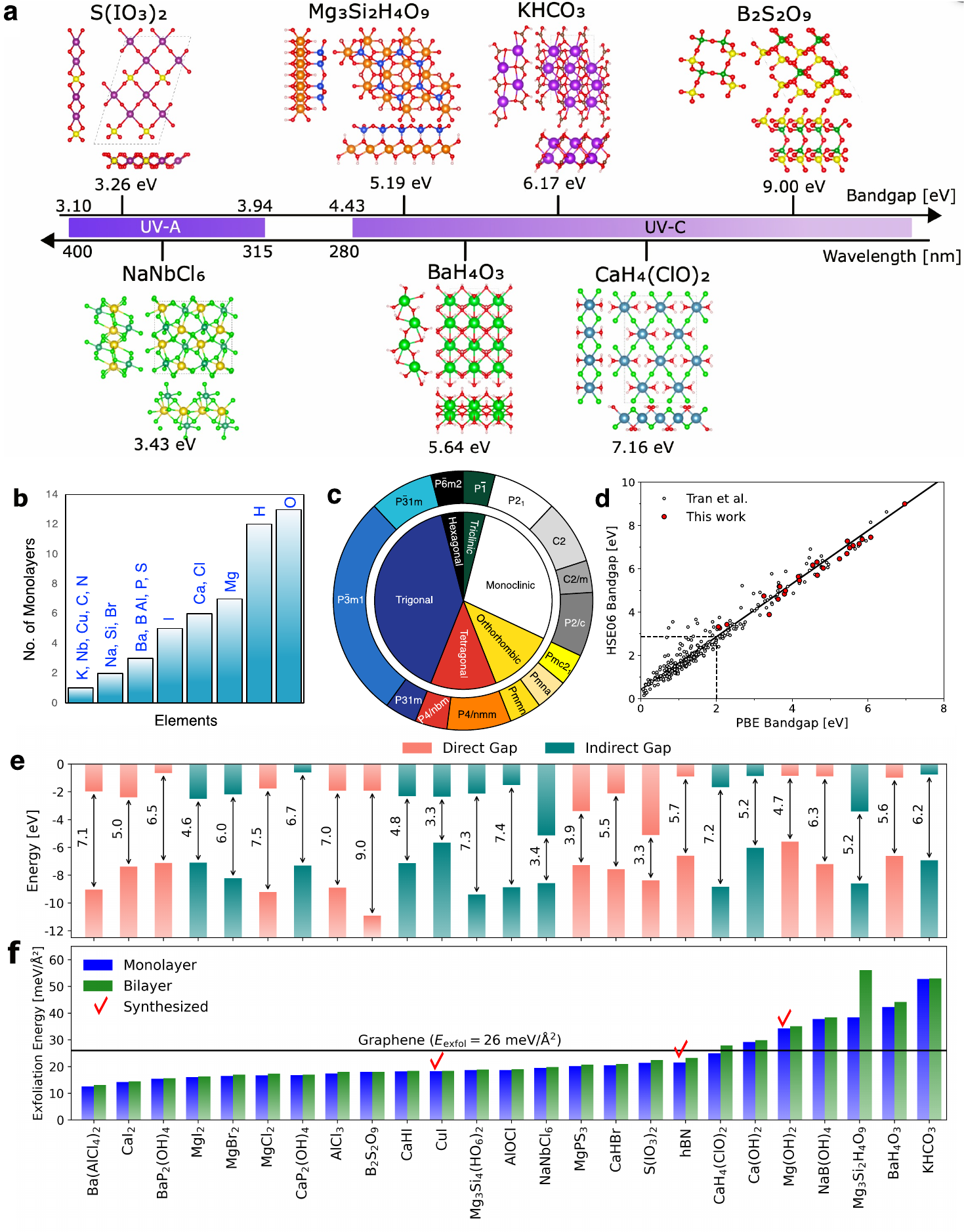}
\caption{\textbf{Overview of the screened candidates.} (a) Illustration of some representative candidate monolayers with bandgaps that lie across the UV-A and UV-C spectrum. (b) Frequency of occurrence of low-risk elements in the candidate monolayers. (c) Distribution of lattice symmetry and spacegroups of the candidate monolayers. (d) Comparison of bandgap values calculated with the HSE06 and PBE functionals, including data from Reference~\onlinecite{monolayer_bandgap_benchmark} and the candidate monolayers from this work. (e) Valence and conduction band edges of the candidate monolayers. The bandgaps (in units of \si{eV}) are presented in the gap region. (f) \textcolor{black}{Exfoliation energies of the monolayer and bilayer forms of the candidate materials. The exfoliation energy of graphene (26~\si{meV/\angstrom^2}) and the monolayer candidates that have previously been synthesized are indicated.}}
\label{UWBG_candidates}
\end{figure*} 

\textbf{Summary of UWBG vdW Materials Screening}. As outlined in Figure~\ref{overview}a, we first apply the \emph{ultrawide bandgap} criteria to the 2,940 `top-down' structures from 2DMatPedia, yielding an initial group of 934 candidates. Further application of the \emph{sustainability and safety} filter on these 934 candidates results in a significantly shrunk pool of 253 candidates. The \emph{synthesizability and stability} filter further reduces the number of candidates to 48. This remaining pool is then pruned for non-vdW entries (those without clear vdW layers and/or with substantial dangling bonds) through manual inspection of their bulk structures. Candidates with more than 25 atoms in the monolayer unit cell are also removed due to their complexity and high computational cost. This results in 30 shortlisted candidates, just 1\% of the initial pool of 2,940 `top-down' 2DMatPedia entries, highlighting the rather stringent nature of this screening framework.

Before subsequent theoretical characterization, preliminary calculations are performed to remove candidate monolayers that exhibit dynamical and elastic instabilities. (i) $\Gamma$-point phonon frequencies obtained with density functional perturbation theory (DFPT) are shown in Tables~S8 and S9. These calculations reveal significant soft modes exceeding 2~\si{meV} in five candidates, which point to dynamical instabilities in their free-standing forms. These five candidates are hence removed. (ii) Ion-relaxed elastic constants computed with the stress-strain method for the remaining 25 candidates are shown in Table~S10. The elastic stiffness tensor in Mandel notation was formed from the elastic constants and diagonalized \citep{elastic_stability_1}. The eigenvalues obtained from this procedure are all positive for these remaining candidate monolayers, indicating their elastic stability \citep{elastic_stability_1, elastic_stability_2}.

As summarized in Figure~\ref{UWBG_candidates}a, the final candidate pool comprises 25 UWBG vdW materials with bandgaps that span the UV-A (315 to 400 \si{nm}) and UV-C (100 to 280 \si{nm}) regime, though none lie in the UV-B regime (280 to 315 \si{nm}). The occurrence frequency of low-risk elements in the final 25 candidates is summarized in Figure~\ref{UWBG_candidates}b. Oxygen and hydrogen are most commonly found, present in nearly half of the shorlisted candidates. Their lattice symmetry and spacegroups are summarized in Figure~\ref{UWBG_candidates}c. The trigonal lattice is the most common among the 25 candidates, followed by the low-symmetry monoclinic lattice. Meanwhile, the P$\bar{3}$m1 spacegroup appears most frequently.

Figure~\ref{UWBG_candidates}d depicts the near-linear empirical mapping between PBE and HSE06 bandgaps of 298 monolayers based on calculations by \citet{monolayer_bandgap_benchmark}, previously used to justify this work's PBE gap criterion. The PBE-HSE06 bandgaps of the 25 screened candidate monolayers from this work are plotted in the same figure, revealing excellent agreement with the empirical trends and validating the PBE gap criterion used. The candidate monolayers' HSE06 band edges with reference to vacuum are shown in Figure~\ref{UWBG_candidates}e.

Even though we used the exfoliation energies provided by 2DMatPedia in the screening procedure, they are in units of \si{meV}/atom while the more commonly used exfoliation energies are in \si{meV/\angstrom}$^2$ \citep{exfoliation}, which requires the energies and surface areas of the layered bulk structure (relaxed with the optB88 functional used in this work for consistency). Having performed these calculations, the monolayer and bilayer exfoliation energies of the candidates are shown in Figure~\ref{UWBG_candidates}f. The exfoliation energies are all below $60$~\si{meV/\angstrom}$^2$, with 19 of the monolayer candidates having exfoliation energies below that of graphene's (26~\si{meV/\angstrom}$^2$ when calculated with the optB88 functional), indicating the ease of their potential synthesis through exfoliation. 

Among the UWBG vdW materials found, several have already been examined more closely for various applications, including \ce{B2S2O9}, \ce{CuI}, \ce{Mg(OH)2}, \ce{Ca(OH)2}, \ce{MgPS3} and h\ce{BN}. Computational studies found that \ce{B2S2O9} possesses strong second-harmonic generation effects and proposed it for applications in deep UV nonlinear optics \citep{B2S2O9_1, B2S2O9_2}. Another previously studied material is \ce{$\beta$-CuI}, whose monolayer has been stabilized under ambient conditions through graphene encapsulation \citep{CuI_expt_1}, and has been found to exhibit promising piezoresistive performance and unexpected room-temperature ferromagnetism (attributed potentially to edge defects) \citep{CuI_expt_2}. \textcolor{black}{These properties of \ce{$\beta$-CuI} (in reduced graphene oxide membranes) motivate its utility in low-power nanosensors and magnetic nanodevices \citep{CuI_expt_2}.} Among the layered hydroxides, monolayer \ce{Mg(OH)2} has been experimentally synthesized and characterized \citep{Mg(OH)2} while \ce{Ca(OH)2} has been exfoliated down to thicknesses ranging between 10 and 100~\si{nm} \citep{Ca(OH)2_1}, with predictions that it can effectively capture \ce{CO2} \citep{Ca(OH)2_2}. \textcolor{black}{The bilayer forms of both layered hydroxides have been theoretically demonstrated to be promising gate dielectrics when paired with monolayer \ce{HfS2} and \ce{WS2} channels for 2D FETs \citep{hydroxides}.} \ce{MgPS3} has been studied theoretically and experimentally for photocatalytic water splitting \citep{MgPS3_dft, MgPS3_ws}. In fact, 21 of the candidate monolayers, including \ce{MgPS3}, have band edges that straddle the water redox potentials (Figure~S\textcolor{black}{13}), making them potential photocatalysts for water-splitting, albeit with UV radiation.  Lastly, the well-characterized h\ce{BN} is one of the final 25 candidates -- a reassuring sanity check for the screening procedure. \textcolor{black}{Known to be the strongest insulating material, with high-temperature stability, low thermal expansion coefficient, high thermal conductivity, corrosion resistance, and rich optoelectronic properties, h\ce{BN} has seen applications as insulating substrates, corrosion-resistant coatings, gate electrics, high-temperature electronics components, and deep UV detectors \citep{hBN_review}.}

\subsection{Piezoelectric Applications}
Seven of the monolayer candidates fall in the piezoelectric crystal classes. Notably, the candidates \ce{NaNbCl6}, \ce{KHCO3} and \ce{S(IO3)2} are similar to h\ce{BN} in that they exhibit piezoelectricity in their monolayer forms but not their bulk forms due to the loss of centrosymmetry going from bulk to monolayer \citep{hBN_piezo_1, hBN_piezo_2}. The piezoelectric stress tensor components $e_{ij}= - d\sigma_j / dE_i$ of these candidates are shown in Table~\ref{piezo}, where $\sigma_j$ are Cauchy stress components in Voigt notation and $E_i$ are electric field components. While the candidates do not exhibit strong out-of-plane piezoelectricity, \ce{KHCO3} and \ce{BaH4O3} have in-plane piezoelectric tensor components that surpass that of h\ce{BN}, hence providing vdW UWBG alternatives other than h\ce{BN} for device applications such as actuators, mechano-electrical sensors, and self-powered UV photodetectors. \textcolor{black}{We further note that six of these seven monolayers (excluding h\ce{BN}) are polar, that is, they possess a spontaneous polarization due to the nonvanishing dipole moment in their unit cell by virtue of their crystal structure, and thus may exhibit pyroelectricity.}

\begin{table}[htb]
\centering
\caption{Piezoelectric stress tensor components (in units of $10^{-10}$~\si{C/m}) of noncentrosymmetric candidate monolayers. Components larger than $1 \times 10^{-10}$~\si{C/m} are in \textbf{boldface}.}
\begin{tabular}{lccccccccc}
\hline\hline
Formula    & $e_{11}$  & $e_{12}$ & $e_{16}$   & $e_{21}$  & $e_{22}$ & $e_{26}$  & $e_{31}$  & $e_{32}$  & $e_{36}$  \\ \hline
\ce{NaNbCl6}  & -0.64         & 0.04          & 0.03           & 0.08          & -0.24          & -0.18 & 0.05  & 0.06 & 0.18  \\
\ce{KHCO3}   & -0.34         & -0.26         & \textbf{1.16}  & -0.22         & \textbf{-4.47} & -0.12 & 0.14  & 0.10 & -0.59 \\
\ce{S(IO3)2}  & 0.42          & 0.05          & -0.23          & -0.07         & -0.02          & -0.18 & 0.00  & 0.00 & 0.16  \\
\ce{B2S2O9}  & \textbf{1.44} & \textbf{1.22} & 0.00           & 0.05          & 0.00           & \textbf{1.03}  & 0.00  & 0.03 & 0.49  \\
\ce{BaH4O3}  & 0.01          & 0.02          & \textbf{-2.29} & 0.12          & \textbf{-2.47} & 0.00  & -0.01 & 0.02 & 0.00  \\
\ce{Mg3Si2H4O9}  & -0.77         & 0.84          & 0.00           & 0.01   & 0.02           & 0.79  & 0.06  & 0.06 & 0.00  \\
h\ce{BN}    & 0.00          & 0.00          & \textbf{1.45}  & \textbf{1.47} & \textbf{-1.47} & -0.05 & 0.00  & 0.00 & 0.00 \\
\hline \hline
\end{tabular}
\label{piezo}
\end{table}

\begin{figure*}[htb]
\centering
\includegraphics[width=\linewidth]{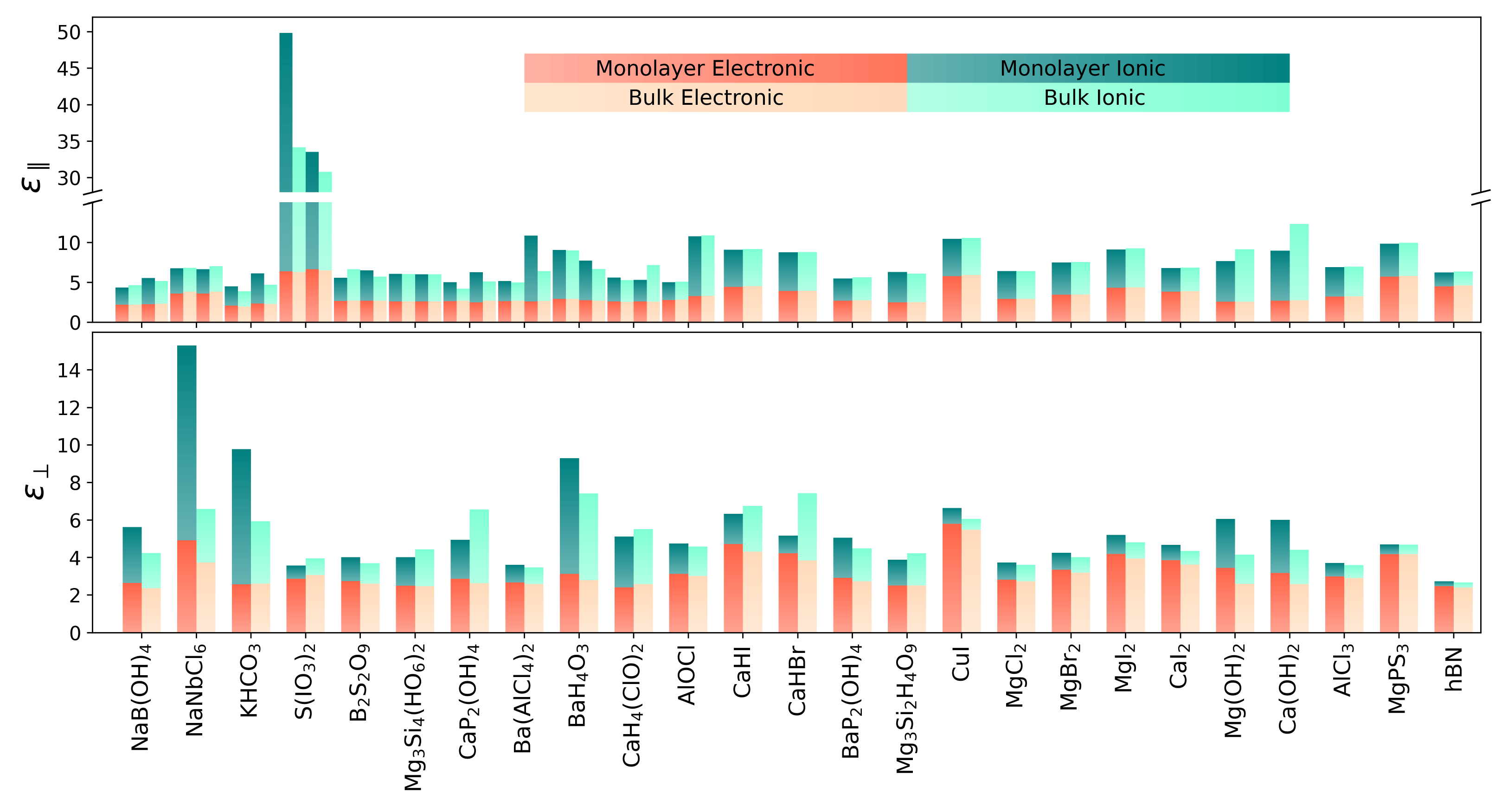}
\caption{\textbf{Static dielectric constants of the candidate materials}. In-plane (upper panel) and out-of-plane (lower panel) dielectric constants of the monolayer and bulk forms of the candidates. The electronic and ionic contributions are indicated by the stacked bars. As the candidates from \ce{NaB(OH)4} to \ce{AlOCl} possess in-plane anisotropy, two sets of in-plane dielectric constants are presented for each candidate.}
\label{dielectric}
\end{figure*} 

\begin{figure*}[htb]
\centering
\includegraphics[width=0.95\linewidth]{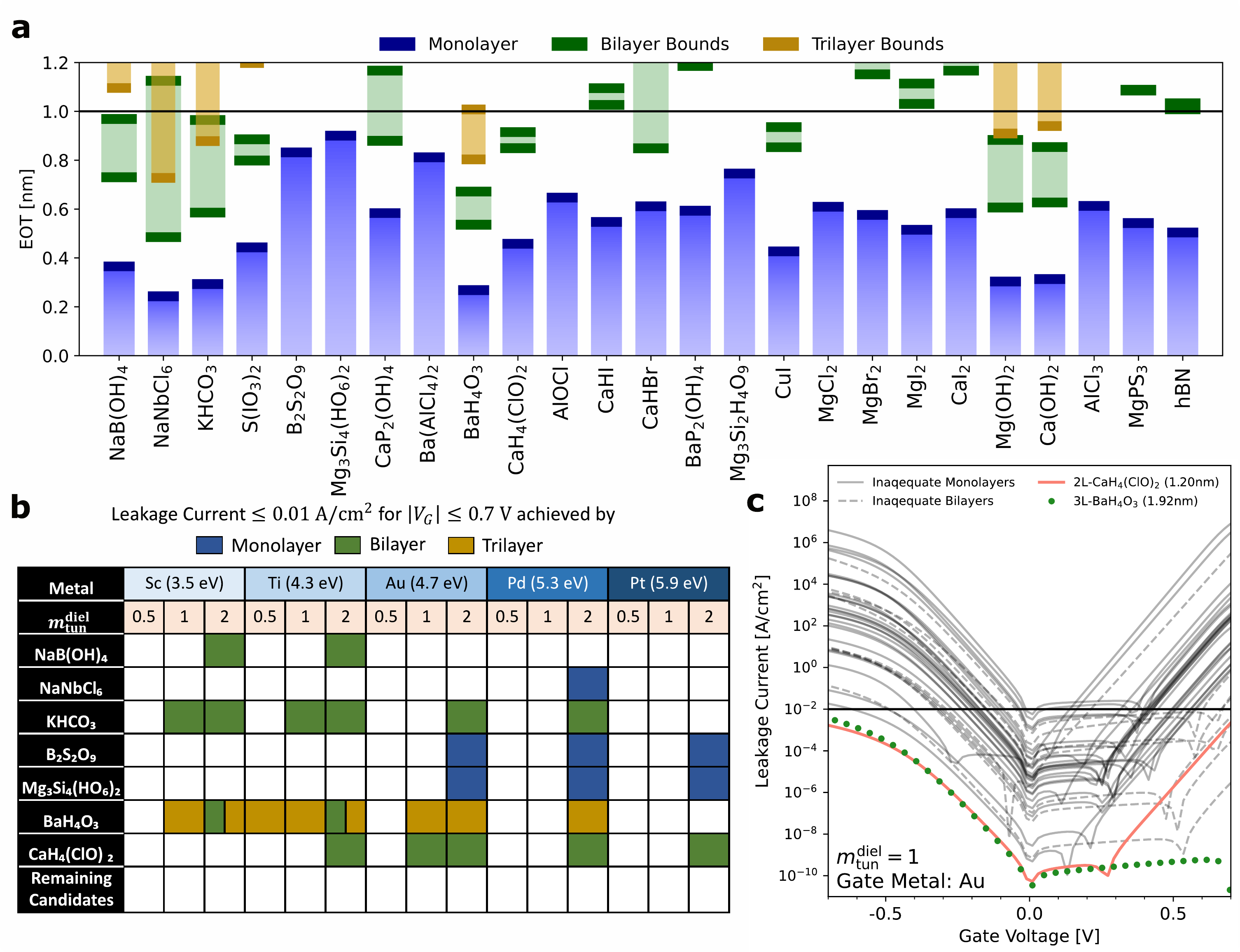}
\caption{\textbf{Ultrawide bandgap candidates for high-$\kappa$ gate dielectric applications.} (a) Equivalent oxide thickness (EOT) of the ultrawide bandgap candidates in their monolayer, bilayer and trilayer forms up to 1.2~\si{nm} EOT. (b) Table indicating the forms of the candidate dielectrics with sub-\si{nm} EOT that falls below the leakage current threshold of 0.01~\si{A/cm^2} for gate voltages $|V_G| \leq 0.7 \text{V}$ when assessed with different gate metal electrodes (work functions indicated beside each metal's name) and tunneling effective masses. (c) Example of a leakage current versus gate voltage plot computed assuming a gold gate electrode and tunneling effective mass of $1m_e$.}
\label{gate_dielectric}
\end{figure*}

\subsection{Dielectric Applications}

The insulating property of UWBG materials makes them viable for dielectric applications. Here we consider the potential roles of 2D UWBG semiconductors in transistor applications as (i) high-$\kappa$ gate dielectrics and (ii) low-$\kappa$ spacers. To assess the dielectric performance of the 2D UWBG candidates, their dielectric constants are computed, postprocessed \citep{hBN_dc_dft, 2d_dielectric_screen} and summarized in Figure~\ref{dielectric}.

We make the following general remarks on the dielectric properties of the UWBG candidates. (i) The electronic contributions to both in-plane and out-of-plane dielectric constants are largely comparable going from the layered bulk to the monolayer limit. The large difference in total dielectric constant between the bulk and monolayer for certain candidates, therefore, can be attributed predominantly to the ionic contribution. (ii) For the in-plane components, \ce{S(IO3)2} possesses the largest dielectric response, ranging from 30 to 35 for its bulk form and up to 50 in its monolayer form. (iii) More pertinent for gate dielectric applications, the monolayer candidates with the largest out-of-plane dielectric constants are \ce{NaNbCl6} (15.3), \ce{KHCO3} (9.8) and \ce{BaH4O3} (8.3). (iv) As the ionic parts of the dielectric constants are significantly larger in some candidates' monolayer forms than in their bulk forms, it is necessary to check if they originate, erroneously, as artefacts from using an insufficiently thick vacuum layer in the DFT simulation cell. To rule out this possibility, the dielectric constants for monolayer \ce{S(IO3)2}, \ce{NaNbCl6}, \ce{KHCO3} and \ce{BaH4O3} are recomputed for cells with varying vacuum thicknesses. The results shown in Table~S15 verifies that they are indeed converged with respect to the vacuum thickness. \textcolor{black}{Additionally, the dielectric constants were recomputed with the finite fields method for comparison against these DFPT-based results using monolayer candidates with very large dielectric constants (\ce{S(IO3)2} and \ce{NaNbCl6}) and h\ce{BN} as test cases. The benchmark data shown in Table S16 demonstrates reasonable agreement of the dielectric constants computed using DFPT and the finite fields method.}

\begin{figure*}[htb]
\centering
\includegraphics[width=0.95\linewidth]{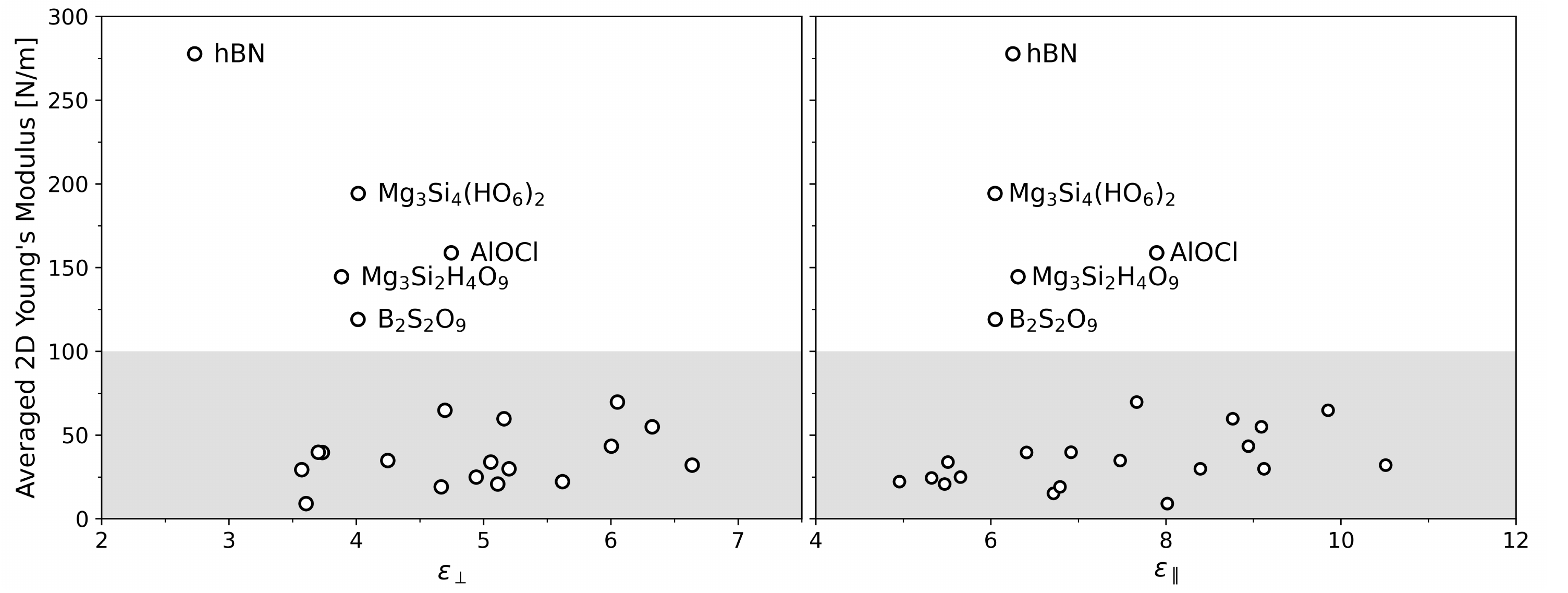}
\caption{\textbf{Ultrawide bandgap candidates for low-$\kappa$ spacer applications.} Plots of angle-averaged 2D Young's moduli against the out-of-plane (left panel) and in-plane (right panel) dielectric constants for the candidate monolayers. Only points corresponding to 2D Young's moduli larger than 100~\si{N/m} are labelled.}
\label{young_diel}
\end{figure*} 

\subsubsection{High-\texorpdfstring{$\kappa$}{k} Gate Dielectrics}

We first consider the role of UWBG vdW semiconductors as high-$\kappa$ gate dielectrics. While conventional \ce{SiO_2} gate dielectrics can be made as thin as 1.2~\si{nm}, the leakage current becomes too high (above 1~\si{A/cm}$^2$ at 1~\si{V} gate voltage), leading to excessive static power dissipation \citep{wong_scaling_2006, robertson_high-k}. The mainstream adoption of \ce{SiO2} dielectrics in silicon-channel transistors arises from the exceptional Si-\ce{SiO2} interface, which can be engineered to achieve low interfacial defect densities \citep{2D_insulators}. In view of this, the absence of dangling bonds on the passivated surfaces of vdW insulators makes them compelling substitutes for \ce{SiO2} \citep{2D_charge_trap}. Although few-layer h\ce{BN} is regarded as a promising ultrathin vdW gate dielectric for transistor miniaturization, sub-\si{nm} equivalent oxide thickness (EOT) h\ce{BN} exhibits unacceptably high leakage currents \citep{hBN_leakage,  2D_insulators, hBN_limits}. The search for performant layered dielectrics \citep{2d_dielectric_screen, TMNH_dielectric, hydroxides, vandenberghe_review} is, however, a challenging quest due to the simultaneous requirements of high $\kappa$, low interfacial defects, high breakdown strength \citep{layer_breakdown} and an appropriate band alignment with the channel semiconductor \citep{dielectric_band_alignment}. Based on the EOT and leakage current requirements, prospective gate dielectrics are identified from the UWBG candidates as follows. \\

\textbf{Equivalent oxide thickness (EOT)}. A key figure of merit of gate dielectric performance, the EOT is
\begin{equation}
    \text{EOT} = \frac{\epsilon_\mathrm{SiO_2}}{\epsilon_{\text{diel}, \perp}}~t_\text{diel},
\end{equation}
for dielectric constant $\epsilon_\text{diel}$, thickness $t_\text{diel}$ and $\epsilon_\mathrm{SiO_2} = 3.9$ \citep{robertson_high-k}. Multilayered structures of the candidates up to the EOT limit of 1~\si{nm} are considered, where the monolayer EOTs are exact while the bi- and trilayer EOTs lie in a range based on the monolayer and bulk dielectric constants. As shown in Figure~\ref{gate_dielectric}a, all monolayer forms of the candidates have sub-\si{nm} EOTs, around a third of them likely have sub-\si{nm} EOT bilayers, and only the trilayer form of \ce{BaH4O3} could possibly have sub-\si{nm} EOT. \\

\textbf{Leakage current}. The leakage current is another important figure of merit for gate dielectrics. The leakage current of sub-\si{nm} EOT hBN was previously calculated in Reference~\onlinecite{hBN_limits} based on a $p$-doped silicon-channel transistor (acceptor doping density $N_A =10^{18}$~\si{cm^{-3}}, donor doping density $N_D =10^{10}$~\si{cm^{-3}}). Here we assess if any of the sub-\si{nm} EOT candidates can outperform hBN under the same set-up, which requires a leakage currents below $0.01$~\si{A/cm^2} for gate voltages $|V_G| \leq 0.7$~\si{V} \citep{hBN_limits}. We estimate the leakage currents based on the Tsu-Esaki formalism \citep{tsu_esaki} implemented in Comphy \citep{comphy_1, comphy_2}, though we note that other tunneling equations, such as the Fowler-Nordheim \citep{fn_model} and Richardson-Dushmann \citep{richardson, dushman} laws, can also be used for the modelling of gate leakage current \citep{2d_dielectric_screen, TMNH_dielectric, hydroxides, vandenberghe_review}.

In the Tsu-Esaki model, the tunneling effective mass $m^\text{diel}_\text{tun}$ is commonly extracted from experimental data in bulk dielectrics based on specific charge transport models \citep{yeo_direct_2000} . For ultrathin vdW materials, however, more care is required in estimating vertical tunneling masses since the vertical effective mass of the layered bulk arises due to the out-of-plane crystal periodicity, a feature that is completely absent in few-layer vdW materials. Moreover, recent experimental work by \citet{tunneling_mass} has demonstrated a peculiar layer-dependent tunneling effective mass for \ce{WSe2} (extracted by fitting the Fowler-Nordheim tunneling equation to their transfer curves). Particularly, the extracted tunneling effective mass of monolayer \ce{WSe2} of 3$m_e$ deviates significantly from the bulk effective mass of 0.2-0.3$m_e$ \citep{tunneling_mass}. While prior studies on ultrathin vdW dielectrics estimate $m^\text{diel}_\text{tun}$ as the vertical effective mass of the corresponding layered bulk structures \citep{2d_dielectric_screen, TMNH_dielectric, hydroxides}, it remains an open question how $m^\text{diel}_\text{tun}$ can be predicted accurately from first-principles. In view of this, we treat $m^\text{diel}_\text{tun}$ as an empirical fitting parameter and consider a range of relative tunneling effective masses (0.5, 1 and 2) in this work. We additionally account for different gate electrode metals with different work functions $\phi$, specifically 
\ce{Sc} ($\phi$=3.5~\si{eV}), \ce{Ti} ($\phi$=4.3~\si{eV}), \ce{Au} ($\phi$=4.7~\si{eV}), \ce{Pd} ($\phi$=5.3~\si{eV}) and \ce{Pt} ($\phi$=5.9~\si{eV}) since the leakage current is also sensitive to the metal/dielectric interfacial barrier heights.

Figure~\ref{gate_dielectric}b shows a summary of whether the sub-\si{nm} EOT candidates met the leakage current criterion under the different parameter combinations while Figure~\ref{gate_dielectric}c presents an example of the variation of leakage current with gate voltage for the candidates when a gold electrode is used and $m^\text{diel}_\text{tun} = m_e$ is assumed.

The leakage current is strongly influenced by the tunneling effective mass and the layer thickness (see Equations~\ref{tsu_esaki_current} and \ref{tunnel_coefficient}). With Pd and Pt gate electrodes, several candidates meet the leakage current criterion when  $m^\text{diel}_\text{tun}=2m_e$, but not for $m^\text{diel}_\text{tun}=m_e$ or  $0.5m_e$. The only candidate that could meet the leakage current criterion, assuming $m^\text{diel}_\text{tun}=0.5m_e$, is trilayer \ce{BaH4O3} (paired with Ti), which can be attributed to its relatively larger thickness of 1.92~\si{nm} compared to other sub-\si{nm} EOT candidates. 
Besides trilayer \ce{BaH4O3}, bilayer \ce{KHCO3} and \ce{CaH4(ClO)2} also meet the leakage current criterion for $m^\text{diel}_\text{tun}=m_e$ (when paired with appropriate gate metals). In summary, the sub-\si{nm} EOT candidates that are most likely to meet the leakage current criterion are trilayer \ce{BaH4O3}, bilayer \ce{KHCO3} and bilayer \ce{CaH4(ClO)2}, when paired with appropriate gate electrodes in a silicon-based FET. 

\textcolor{black}{Additional thermal stability assessments were performed using molecular dynamics simulations for the notable candidates including \ce{S(IO3)2} (large in-plane dielectric response) and the promising gate dielectric candidates \ce{BaH4O3}, \ce{KHCO3} and \ce{CaH4(ClO)2} (see Figures~S4 to S10). The candidates were found to be thermally stable at 300~\si{K}, with the exception of \ce{KHCO3}, which wrinkles instead. Furthermore, molecular dynamics simulations were employed to assess the stability of the gate dielectric candidates \ce{BaH4O3}, \ce{KHCO3} and \ce{CaH4(ClO)2} under high temperature (500~\si{K}) and moist conditions at 300~\si{K} (simulated with an adsorbed water molecule on the surface of the structures). \ce{BaH4O3} was found to be stable at both high temperature moist conditions, \ce{KHCO3} exhibited similar wrinkling behaviour as before, while \ce{CaH4(ClO)2} underwent structural changes under these conditions (an \ce{OH} molecule was dislodged from the structure at 500~\si{K} while an \ce{HCl} molecule was dislodged instead under moist conditions). Overall, \ce{BaH4O3} emerges as the most favorable gate dielectric candidate for its sub-\si{nm} EOT, low leakage currents and stability under high temperature and moist conditions.}

\textcolor{black}{While these results hold for the pristine forms of the candidates, the presence of defects can diminish device performance (for example, trap-assisted tunneling can raise the leakage current). Consideration of defects, however, requires knowledge of specific experimental conditions during material synthesis and device fabrication, and is thus beyond the scope of this work. The results presented here are thus the `best case scenario' estimates, similar to that of \citet{hBN_limits} which demonstrated that even in the pristine limit, h\ce{BN} would not meet the requirements of a gate dielectric. Importantly, the computational screening performed in this work identifies alternative layered material candidates that may outperform h\ce{BN} in the pristine limit, thus re-establishing the potential importance of layered materials in gate dielectric applications.}

\begin{figure*}[htb]
\centering
\includegraphics[width=\linewidth]{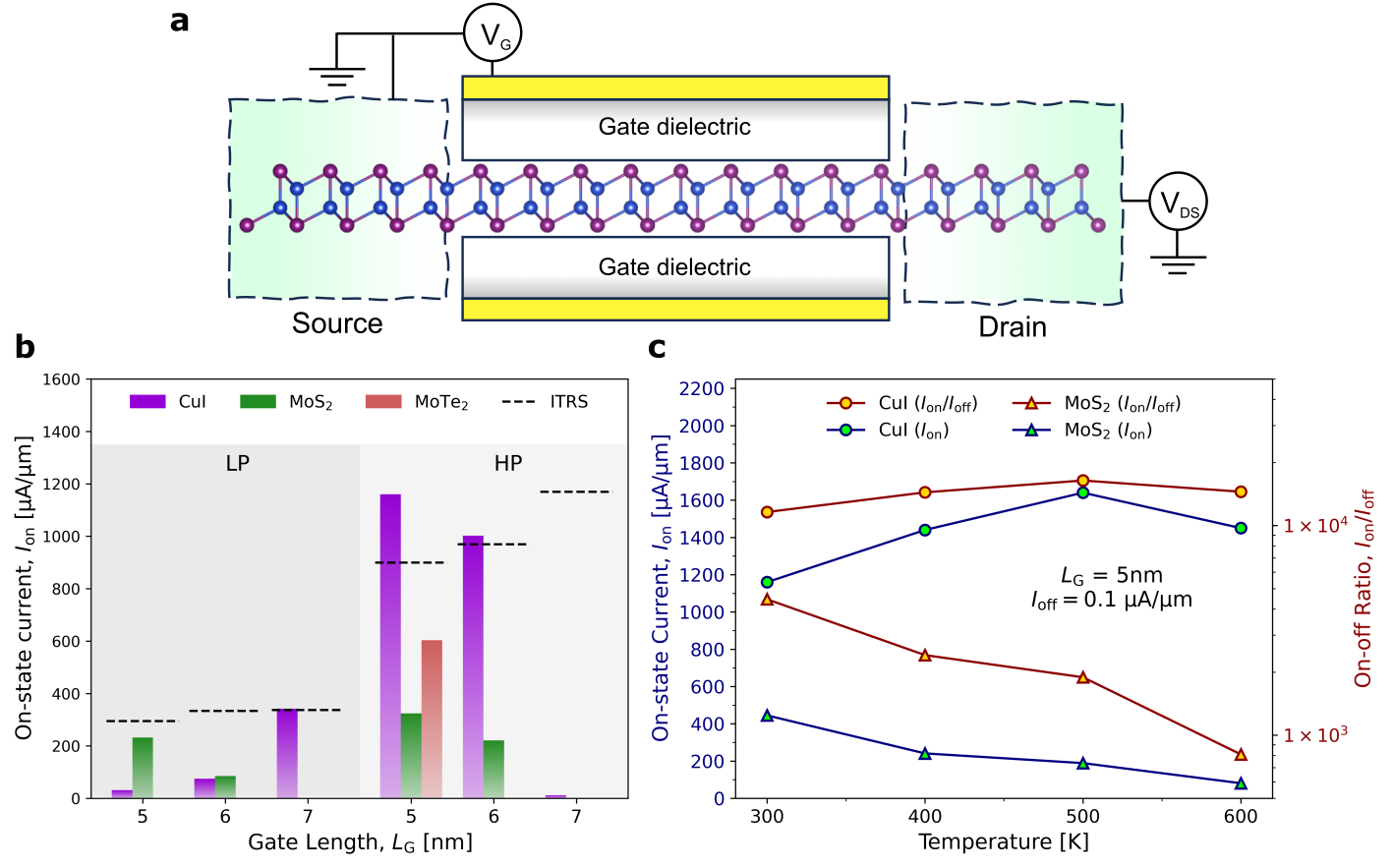}
\caption{\textbf{Monolayer \ce{CuI} nanochannel device performance.} (a) Schematic of a double-gated metal-oxide-semiconductor field-effect transistor (MOSFET) with a monolayer \ce{CuI} channel. (b) On-state current of $5$, $6$ and $7$~\si{nm} gate length devices, benchmarked against International Technology Roadmap for Semiconductors (ITRS) requirements. Monolayer \ce{MoS2} and \ce{MoTe2} MOSFET data from literature  \citep{transistor_benchmark_1, transistor_benchmark_2, transistor_benchmark_3} are also presented. (c) Temperature dependence of the on-state current and current on-off ratio for \ce{CuI} and \ce{MoS2} MOSFETs at 5~\si{nm} gate lengths.}
\label{power_electronics}
\end{figure*}

\subsubsection{Low-\texorpdfstring{$\kappa$}{k} Spacers}

Low-$\kappa$ dielectrics are useful as transistor spacers designed to diminish the parasitic capacitance effects between metal electrodes, which are particularly severe in aggressively scaled transistor architecture. This parasitic effect can exacerbate the resistance–capacitance delay and degrade device- and system-level performance \citep{dielectrics_2Dtransistor}. For front-end-of-line (FEOL) transistor technologies, the spacers are positioned between the source/drain metal and the gate/channel stacks in a FinFET or NS FET; for back-end-of-line (BEOL) interconnect technologies, the spacers are placed between the interconnect layers \citep{dielectrics_2Dtransistor}. For low-$\kappa$ spacer applications, mechanical strength is an important consideration due to reliability issues during device fabrication \citep{dielectrics_2Dtransistor}. 
In Figure~\ref{young_diel}, angle-averaged 2D Young's moduli of the candidates monolayers are computed and plotted against their dielectric constants. We identify \ce{hBN}, \ce{Mg3Si4(HO6)2}, \ce{Mg3Si2H4O9}, \ce{B2S2O9} and \ce{AlOCl} as potential candidates for low-$\kappa$ layered spacer applications due to their  low-$\kappa$ and high 2D Young's modulus (above 100~\si{N/m}), with hBN possessing the largest 2D Young's modulus of 278~\si{N/m} along with the smallest (in-plane) dielectric constant of 2.7.

\subsection{Nanochannel Transistor for Computing Electronics}

2D semiconductors are promising candidates in continuing transistor scaling beyond the 12-nm gate-length limit of conventional silicon devices \citep{2D_FET, chip1}. UWBG 2D semiconductors, such as metal oxides \citep{TeO2_1, TeO2_2, TeO2_3}, have been actively explored for low-power transistor applications since their larger bandgap is beneficial for suppressing the off-state current. 

To illustrate the potential of UWBG vdW materials for low-power and high-performance transistor applications, we perform in-depth DFT+NEGF quantum transport calculations on the monolayer form of $\beta$-\ce{CuI}. This choice of \ce{CuI} is motivated by several reasons. Chiefly, monolayer \ce{CuI} has been experimentally synthesized at ambient conditions \citep{CuI_expt_1} and has been shown to be stable up to 900~\si{K} \citep{CuI_expt_2} \textcolor{black}{(in good agreement with its observed stability at 300~\si{K} and 500~\si{K} based on molecular dynamics simulations in Figure~S11)}. Furthermore, monolayer \ce{CuI} has a `sweet spot' bandgap of 3.3~\si{eV}, large enough to suppress the off-state current generated by the Boltzmann tail, $I_\text{off} \propto \exp{(- E_g / \alpha k_B T)}$ \citep{onoffcurrent_gap}, but also small enough to be compatible with most gate dielectrics with larger bandgaps.

The device model used comprises a double-gated metal-oxide-semiconductor field-effect transistor (MOSFET) with a \ce{CuI} channel (Figure~\ref{power_electronics}a). We calculate the transport characteristics of \ce{CuI} MOSFETs for nanochannel devices with gate lengths $L_\text{G}=5,~6$ and $7$~\si{nm}. Only ballistic transport, which dominates over electron-phonon scattering in the sub-10~\si{nm} gate length regime \citep{liu2013monolayer}, is considered. The on-state current, $I_\text{on}$, a key figure of merit for transistor performance, is extracted and benchmarked in accordance to International Technology Roadmap for Semiconductors (ITRS) standards \citep{ITRS}, alongside performance metrics of monolayer \ce{MoS2} and \ce{MoTe2} MOSFETs from other studies \citep{transistor_benchmark_1, transistor_benchmark_2, transistor_benchmark_3}. As shown in Figure~\ref{power_electronics}b,
the $L_\text{G}=7$~\si{nm} \ce{CuI} MOSFET fulfills the ITRS requirements for low-power (LP) devices while the requirements for high-performance (HP) electronics are satisfied at $L_\text{G}=5$ and $6$~\si{nm}. Meanwhile, none of the $L_\text{G}=5,~6$ and $7$~\si{nm} \ce{MoS2} and \ce{MoTe2} MOSFETs from other theoretical studies meet the LP and HP ITRS criteria. Notably, \ce{CuI} achieves both \emph{materials-level} sustainability and \emph{device-level} sustainability in conforming to the \emph{sustainability and safety} filter proposed in this work, in addition to the ITRS standards for sub-10~\si{nm} computing electronics.

Beyond ambient condition operations, UWBG materials such as \ce{SiC} and \ce{GaN} are often used in high-temperature applications common in automotive, aerospace and deep-well drilling industries \citep{high_temp}. These applications stand to benefit from more compact and robust electronic circuits, especially for situations where space constraints are paramount \citep{power_electronics}. Miniaturization, however, exacerbates thermal issues since power densities have increased \citep{power_electronics}. UWBG vdW materials are hence especially apt as their ultrathin, few-layered nature enables the fabrication of smaller devices while their large bandgaps contribute to the maintenance of electronic performance under high-temperature conditions.

Monolayer \ce{CuI}, whose bandgap of 3.3~\si{eV} is comparable to that of the common power electronics material \ce{GaN} (3.4~\si{eV}) \citep{GaN_gap}, is thus potentially suited for high-temperature transistor operation. Transport simulations for \ce{CuI} and \ce{MoS2} devices are performed for temperatures ranging from 300~\si{K} to 600~\si{K}. We fix $L_\text{G}=5$~\si{nm} and set the off-state current, $I_\text{off}$, as 0.1 \si{\mu A/ \mu m}. The temperature variation of the on-state current, $I_\text{on}$, and current on-off ratio, $I_\text{on}/I_\text{off}$, extracted from the transport simulations are shown in Figure~\ref{power_electronics}c. The \ce{CuI} MOSFET's $I_\text{on}$ and $I_\text{on}/I_\text{off}$ increases from 300 to 500~\si{K} and decreases slightly from 500 to 600~\si{K}. In contrast, these two figures of merit monotonically worsen as the temperature is raised for the \ce{MoS2} MOSFET, which is consistently outperformed by the \ce{CuI} MOSFET over the temperature range by rather large margins. The preservation of nanochannel device performance under high temperatures makes \ce{CuI} a prospective channel material in power electronics applications, and offers a viable route towards the miniaturization of computing electronics for extreme environments.

\begin{figure*}[htb]
\centering
\includegraphics[width=0.95\linewidth]{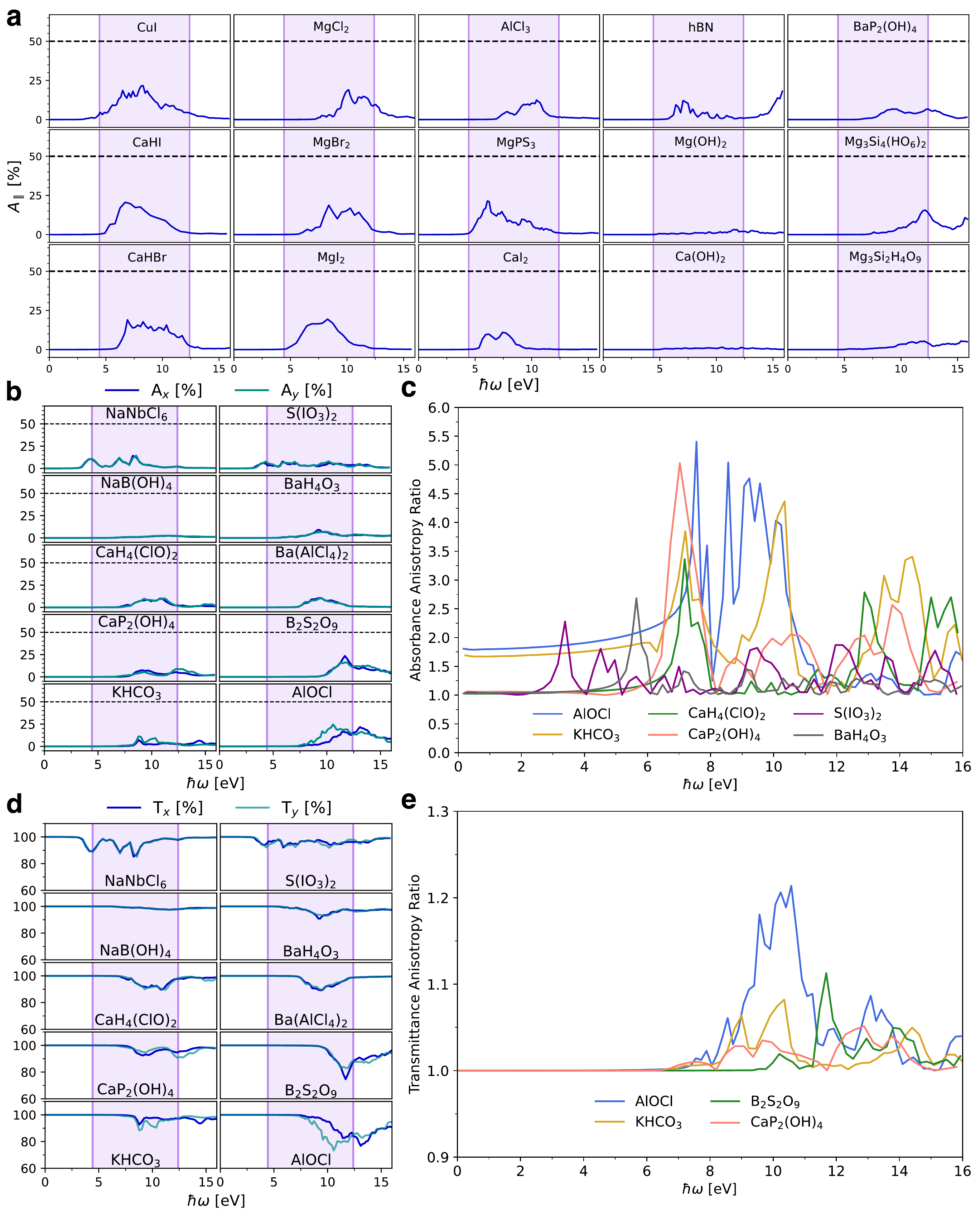}
\caption{\textbf{Ultrawide bandgap candidates for ultraviolet optoelectronics.} (a) Absorbance percentages of the isotropic candidates. (b) Absorbance percentages, (c) absorbance anisotropy ratios , (d)  transmittance percentages and (e) transmittance anisotropy ratios of the anisotropic candidates when light is normally incident on their monolayer forms. The maximum absorbance percentage of 50\% is indicated by a dotted horizontal line in (a) and (b) while the ultraviolet-C (4.43 to 12.4~\si{eV}) regime is shaded in (a), (b) and (d).}
\label{optical_spectra}
\end{figure*}

\subsection{Ultraviolet Optoelectronics}

We now examine the viability of the UWBG vdW candidates in ultraviolet (UV) optoelectronics. UV photodetectors are widely used in applications that range from flame sensors \citep{UV_flame}, dermatological imaging devices \citep{UV_dermatology} and astronomical instruments \citep{UV_astro} to military technology such as missile warning systems \citep{UV_missile} and detectors for chemical and biological warfare agents \citep{UV_chemical, UV_biological}. Often, it is paramount for such photodetectors to be solar-blind, that is, they only absorb UV-C radiation (which is entirely absorbed by the ozonosphere and should not exist naturally within the Earth's atmosphere), so as to minimize interference from background solar radiation. Optically anisotropic UWBG materials can additionally serve as polarization-sensitive photodetectors or polarizers \citep{meng_terahertz_2022}.

To characterize the optical properties of the UWBG vdW candidates, HSE06 dielectric functions are computed and postprocessed into frequency-dependent transmittance (T), absorbance (A) and reflectance (R) percentages for the idealized set-up where light is normally incident on the candidate monolayers suspended in air \citep{dielectric_function_1, dielectric_function_2}. We first note that 21 of the 25 candidates (the exceptions being \ce{S(IO3)2}, \ce{CuI}, \ce{NaNbCl6} and \ce{MgPS3}) have bandgaps larger than 4.43~\si{eV}, the minimum energy of a UV-C photon, and can hence be used as solar-blind UV photodetectors. This is further illustrated in the absorbance plots in Figures~\ref{optical_spectra}a and b, where these 21 candidates have near-zero percent absorbance for photon frequencies below the lower bound of the UV-C frequency regime, rendering them insensitive towards the background solar radiation present in the atmosphere. \textcolor{black}{Moreover, of these 21 candidates, we observe broadband absorbance of 15\% for several candidates, including \ce{CaHI}, \ce{CaHBr}, \ce{MgBr2} and \ce{MgI2} (note that the theoretical maximum absorbance percentage is 50\% \citep{dielectric_function_2})}.

For polarization-sensitive UV photodetection applications, we consider the absorbance anisotropy ratio of the anisotropic candidates, given by $\text{max}(A_x, A_y)/\text{min}(A_x, A_y)$. Figures~\ref{optical_spectra}b and c support \ce{AlOCl} as a viable candidate for such applications, given its relatively large absorbance anisotropy ratio reaching up to \textcolor{black}{5} and the fact that absorbance in one principal direction is consistently larger than the absorbance in the other direction in the UV-C regime.
We perform a similar analysis to assess the viability of the anisotropic candidates as UV-C polarizers. The transmittance percentages and transmittance anisotropy ratio, given by $\text{max}(T_x, T_y)/\text{min}(T_x, T_y)$, of the candidates are shown in Figures~\ref{optical_spectra}d and e. Once again, \ce{AlOCl} is the most viable candidate, with transmittance anisotropy ratios \textcolor{black}{up to 1.2} in the UV-C regime. \textcolor{black}{\ce{AlOCl} is also thermally stable at 300~\si{K} based on molecular dynamics simulations (Figure~S12).}

\begin{figure*}[htb]
\centering
\includegraphics[width=0.95\linewidth]{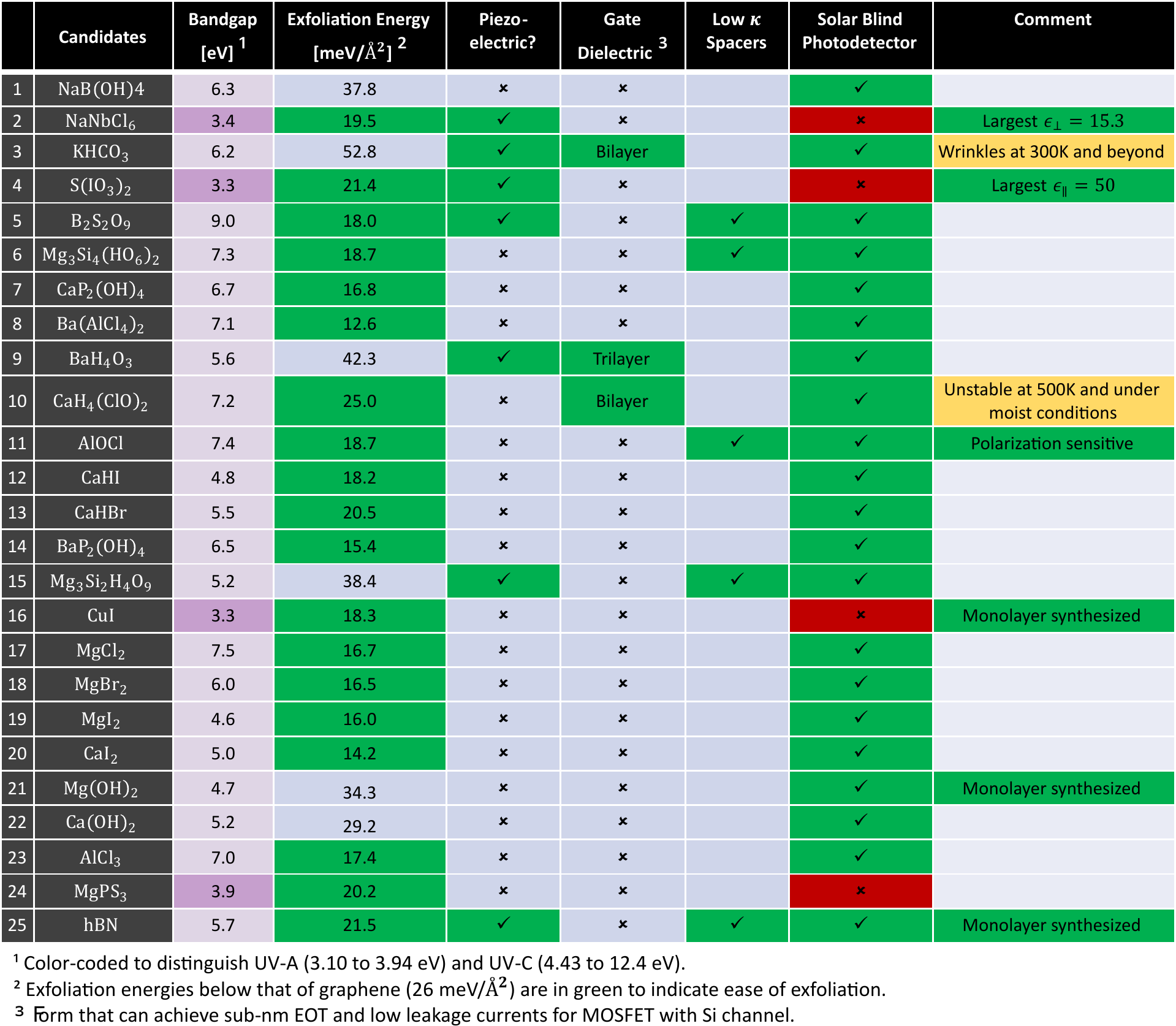}
\caption{\textbf{Summary of the 25 computationally-screened candidates.} The electronic and mechanical properties, as well as the potential (opto)electronic device applications of the 25 UWBG candidates are listed.}
\label{summary}
\end{figure*}

\section{\label{conclusion} Conclusion}

In summary, we performed a sustainability-guided search for UWBG vdW materials. Of the 2,940 `top-down' entries from 2DMatPedia considered, we arrived at a modest candidate pool of 25 sustainable UWBG vdW materials. \textcolor{black}{Trilayer \ce{BaH4O3} was identified to meet the requirements for gate dielectric applications in a silicon channel transistor}, while monolayer \ce{hBN}, \ce{Mg3Si4(HO6)2}, \ce{Mg3Si2H4O9}, \ce{B2S2O9} and \ce{AlOCl} were proposed as mechanically strong low-$\kappa$ spacers. Monolayer \ce{CuI} was found to be a promising channel material for sub-10~\si{nm} FETs that met the ITRS requirements at room temperature and at high temperatures up to 600~\si{K}, indicating strong potential in power electronics miniaturization. Lastly, most of the candidates had excellent solar-blind UV absorbance while \ce{AlOCl} was the most promising candidate for polarization-sensitive UV photodetection and polarizer applications.

In spite of the stringent requirements for sustainability and safety in this work's screening efforts, it remains possible to obtain candidates that are promising (albeit not as exceptional as in other works) for various (opto)electronics applications. To address the central dilemma of this work -- the trade-off between device performance and sustainability concerns -- our findings demonstrate that it is indeed possible to maintain a balance between achieving acceptable device performance and maintaining materials-level sustainability in such computational screening efforts.

Having addressed \emph{materials-level} sustainability in this work, we note that \emph{process-level} sustainability is also a pressing concern, which encompasses aspects of materials synthesis and device fabrication in terms of the required reagents and byproducts formed. It remains an open question whether \emph{process-level} considerations can be accounted for during the materials screening process, providing an avenue of further work to imbue the materials discovery process with an even greater focus on sustainability.

\section{\label{methods} Methods}

\subsection{General DFT Calculations}

The Vienna Ab initio Simulation Package (VASP) \citep{VASP_1, VASP_2} was employed, with a plane-wave cutoff of 520 \si{eV} and Brillouin zone sampling frequency of $0.03 \times 2\pi$ \si{\angstrom^{-1}} used for all DFT calculations. The optB88 van der Waals exchange-correlation (XC) functional \citep{optB88_1, optB88_2, optB88_3, optB88_4} was used as an affordable method for structural calculations that can reproduce exfoliation energy trends and interlayer interactions reasonably close to the more accurate but computationally costly random-phase approximation (RPA) method \citep{optB88_RPA_1, optB88_RPA_2}. Geometry relaxations were performed such that forces and stresses were converged below $0.002$~\si{eV/\angstrom} and $0.002$~\si{kBar} respectively for the monolayers, bilayers and bulk structures of the candidate materials. \textcolor{black}{Molecular dynamics simulations were performed with the PBE-D3 functional \citep{PBE, D3}, the Nose-Hoover thermostat and timesteps of 1.5~\si{fs}, with plane-wave cutoffs of 520 \si{eV}, Brillouin zone sampling frequencies of $0.06 \times 2\pi$ \si{\angstrom^{-1}} (which led $\Gamma$-point sampling for the supercells used for the simulations).} Band structure calculations were performed using the HSE06 functional \citep{HSE06}. We consider HSE06 bandgaps (computed on structures optimized with the optB88 vdW functional) to be an affordable and sufficiently accurate estimate for experimental gaps based on our HSE06@optB88-versus-experimental monolayer bandgap benchmark in Table~S2.

\subsection{Dielectric Calculations}

The dielectric constants (including both electronic and ionic contributions) of the candidate monolayers and their layered bulk parents were computed using density functional perturbation theory (DFPT) with the optB88 functional, including local field effects in DFT, \textcolor{black}{allowing us to obtain the dielectric constants directly without explicit calculation of the electric polarization}. Due to the periodic boundary conditions used in plane-wave DFT codes, DFT simulations for slabs require the construction of slab-vacuum supercells. It is hence necessary to postprocess the supercell dielectric constants, $\epsilon_\text{SC}$, obtained directly from VASP into monolayer dielectric constants, $\epsilon_\text{mono}$. The relationship for the out-of-plane component is
\begin{equation}
    \epsilon^\perp_\text{mono} = \left[1 + \frac{L}{t} \left(\frac{1}{\epsilon^\perp_\text{SC}} - 1\right) \right]^{-1},
\end{equation}
while the in-plane components obey
\begin{equation}
    \epsilon^\parallel_\text{mono} = 1 + \frac{L}{t} \left(\epsilon^\parallel_\text{SC} - 1\right),
\end{equation}
where $L$ is the supercell height and $t$ is the monolayer thickness, obtained as the interlayer distance of their corresponding bilayers \citep{hBN_dc_dft, 2d_dielectric_screen}. For the 25 candidates in this work, we note that there is broad agreement (less than $5\%$ difference) of the monolayer thicknesses (extracted from the bilayer geometry) and the thickness per layer extracted from the bulk as shown in Figure~S3. While this suggests that it may be permissible to extract monolayer thicknesses directly from the bulk geometry and avoid relaxing the bilayers, a more thorough examination is recommended.

Comphy \citep{comphy_1, comphy_2} was used for the leakage current calculations based on the Tsu-Esaki formalism \citep{tsu_esaki}. Within this formalism, the tunneling current density between two carrier reservoirs separated by an insulating barrier is expressed as
\begin{equation}
\label{tsu_esaki_current}
    J_\text{tunnel} = \frac{m^\text{ch}_\text{eff} q_0}{2\pi^2\hbar^3} \int_{E_\text{CB}}^\infty \Theta_e(E) N_e(E) dE,
\end{equation}
where $m^\text{ch}_\text{eff}$ is the effective electron mass in the semiconductor channel, $E_\text{CB}$ is the conduction band energy,
\begin{equation}
    N_e(E) = k_B T \ln \left( \frac{1+\exp\left(-\frac{E-E_{\text{F},1}}{k_BT}\right)}{1+\exp\left(-\frac{E-E_{\text{F},2}}{k_BT}\right)} \right)
\end{equation}
is the electron supply function for temperature $T$ and reservoir Fermi levels $E_{\text{F},1(2)}$, and
\begin{equation}
\label{tunnel_coefficient}
    \Theta_e(E) = \exp \left[-\frac{4d\sqrt{2m^\text{diel}_\text{tun}}}{3\hbar q_0 V_G} \left[ (q_0\phi - E)^\frac{3}{2} - (q_0\phi_0 - E)^\frac{3}{2} \right] \right]
\end{equation}
is the carrier tunneling coefficient for layer thickness $d$, applied voltage $V_G$, energy barrier heights $(\phi,~\phi_0)$ and the tunneling effective mass $m^\text{diel}_\text{tun}$. $q_0$ is the elementary charge, $\hbar$ is the reduced Planck's constant and $k_B$ is Boltzmann's constant.  \\

The Young's modulus was obtained from the elastic constants (computed with the stress-strain method) using the expression
\begin{equation}
    E(\theta) = \frac{d_0}{C_{11}\sin^4\theta + C_{22}\cos^4\theta + (\frac{d_0}{C_{66}}-2C_{12}) \sin^2\theta\cos^2\theta}
\end{equation}
for angle $\theta$ and $d_0 = C_{11}C_{22}-C_{12}^2$ \citep{youngs_modulus}. A further step to compute the angle-averaged 2D Young's modulus, $E=\int_0^{2\pi} E(\theta) d\theta$, was taken to simplify its representation in Figure~\ref{young_diel}. In the same vein, the in-plane dielectric constants in Figure~\ref{young_diel} were averaged for anisotropic candidates such that $\epsilon_\parallel = (\epsilon_{11}+\epsilon_{22})/2$.

\subsection{Atomistic Device Simulations}

The DFT+NEGF quantum transport simulations were performed with QuantumATK 2022 \citep{quantumATK}. The Landauer–B\"uttiker formula,
\begin{equation}
    I_\text{DS} = \frac{2e}{h} \int_{-\infty}^\infty \left[f_\text{D}(E-\mu_\text{D}) - f_\text{S}(E-\mu_\text{S}) \right] T(E) dE,
\end{equation}
was adopted to calculate the drain current $I_\text{DS}$, where $f_\text{S}~(f_\text{D})$ is the Fermi-Dirac distribution function of source (drain), $\mu_\text{S}~(\mu_\text{D})$ is the chemical potential of source (drain), and $T(E)$ represents the transmission function. $T(E)$ is averaged over the $\mathbf{k}$-dependent transmission coefficients $T_\mathbf{k}(E)$ in the irreducible Brillouin zone, and computed as
\begin{equation}
    T_\mathbf{k}(E) = Tr[T^S_\mathbf{k}(E)G_\mathbf{k}(E) T^D_\mathbf{k}(E) G^\dagger_\mathbf{k}(E)],
\end{equation}
where $G_\mathbf{k}(E)$ ($G^\dagger_\mathbf{k}(E)$) represents the retarded (advanced) Green’s function and $T^{S/D}_\mathbf{k}(E) = i[\Sigma_\mathbf{k}^{S/D} - (\Sigma_\mathbf{k}^{S/D})^\dagger]$ is the energy level broadening for self-energy $\Sigma_\mathbf{k}^{S/D}$, which represents the coupling between source/drain electrodes and channel. 

A k-point mesh of $7 \times 1 \times 132$ (for both the electrode and channel), a density mesh cutoff of 105~Hartree and PseudoDojo pseudopotentials were used for the device simulations. Periodic, Neumann, and Dirichlet boundary conditions were set for the transverse, vertical, and transport directions, respectively. The source and drain electrodes were simulated implicitly through doping. Systematic testing of various doping concentrations yielded an optimal value of $1\times 10^{13}$~\si{cm^{-2}}.

\subsection{Optical Properties Calculations}

The frequency-dependent dielectric functions were computed within the independent particle approximation at the HSE06 hybrid level \citep{HSE06}. \textcolor{black}{While the GW-BSE method can capture excitonic effects beyond an independent-particle approach, its prohibitive computational cost, exacerbated by the difficulty of converging the calculation with respect to several interdependent convergence parameters, renders it an unfavourable tool for screening the candidates \citep{gwbse}. In spite of this, our focus lies on the more qualitative aspects of the optical spectra whereby the use of the independent particle approximation may be sufficient.} To postprocess the in-plane dielectric functions into frequency-dependent transmittance (T), absorbance (A) and reflectance (R) percentages, the in-plane supercell dielectric function $\epsilon_\text{SC}^\parallel(\omega)$ obtained directly from VASP is first converted into the in-plane 2D conductivity \citep{dielectric_function_1},
\begin{equation}
    \sigma_\text{2D}^\parallel(\omega) = -i\epsilon_0 \omega L \left[\epsilon_\text{SC}^\parallel(\omega) - 1 \right],
\end{equation}
where $\epsilon_0$ is the vacuum permittivity and $\omega$ are the photon frequencies. For incoming light at normal incidence onto the monolayers suspended in air, the TAR \citep{dielectric_function_1, dielectric_function_2} are then computed as
\begin{equation}
    \text{T} =  \frac{1}{\left|1 + \tilde{\sigma}/2\right|^2}  ~,~ 
    \text{A} =  \frac{\mathbb{R}\text{e}\{\tilde{\sigma}\}}{\left|1 + \tilde{\sigma}/2\right|^2} 
    ~,~ \text{R} = \left| \frac{\tilde{\sigma}/2}{1 + \tilde{\sigma}/2} \right|^2,
\end{equation}
where $\tilde{\sigma}(\omega) = \sigma_\text{2D}^\parallel(\omega)/\epsilon_0 c$ for speed of light $c$. \\

\begin{acknowledgments}
\textcolor{black}{The authors would like to thank Stefano Falletta for helpful discussions on the computation of dielectric constants within density functional theory.}

\texttt{pymatgen} \citep{pymatgen} served as an invaluable tool for pre- and postprocessing the VASP calculations, while VESTA \citep{VESTA} was crucial for structure vizualization.

This work is supported by the Singapore Ministry of Education Academic Research Fund Tier 2 (Award No. MOE-T2EP50221-0019), SUTD Kickstarter Initiatives (SKI) (Award No. SKI 2021\_01\_12) and the SUTD-ZJU IDEA Thematic Research Grant Exploratory Project (SUTD-ZJU (TR) 202203). L.X. is supported by the China Scholarship Council and the Fundamental Research Funds for the Central Universities. J.L. is supported by the Ministry of Science and Technology of China (No. 2022YFA1203904), the National Natural Science Foundation of China (No. 91964101, and No. 12274002).
The computational work for this article was performed on resources of the National Supercomputing Centre, Singapore (https://www.nscc.sg), the High-Performance Computing Platform of Peking University \textcolor{black}{and the computational resources provided by the FAS Division of Science Research Computing Group at Harvard University.}
\end{acknowledgments}

\section*{Conflict of Interest}
The authors declare no conflict of interest.

\section*{Author Contributions}
C.W.T. and L.X. contributed equally to this work. C.W.T. performed conceptualization (equal); data curation (lead); formal analysis (lead); visualization (equal); wrote original draft (lead); wrote review and editing (equal). L.X. performed data curation (equal); formal analysis (supporting); wrote original draft (supporting). C.C.E. performed data curation (equal). S.‐P.C. and B.K. acquired resources (equal). H.Y.Y.performed funding acquisition (equal). S.A.Y. performed supervision (supporting). J.L. acquired resources (equal); supervision (supporting). Y.S.A. performed conceptualization (equal); funding acquisition (equal); acquired resources (equal); supervision (lead); visualization (equal); wrote original draft (supporting); wrote review and editing (equal).

\section*{Data Availability Statement}
The data that support the findings of this study are available from the corresponding author upon reasonable request.


%

\end{document}